# Sample size determination in superiority or non-inferiority clinical trials with time-to-event data under exponential, Weibull and Gompertz distributions

**Dong Han**[1,2], Yawen Hou[3], Zheng Chen[1*],

1: Department of Biostatistics , School of Public Health (Guangdong Provincial Key Laboratory of Tropical Disease Research), Southern Medical University, No.1023, South Shatai Road, Baiyun District, Guangzhou 510515, China;  2: Department of Quality Control, The Third Affiliated Hospital of Southern Medical University, Guangzhou 510665, China;  3: Department of Statistics, College of Economics, Jinan University, Guangzhou 510632, China

* Corresponding author: Zheng Chen

**Abstract:** To examine the effect of exponential, Weibull and Gompertz distributions on sample size determination for superiority trials (STs) or non-inferiority trials (NTs) with time-to-event data, we present two sample size formulas for STs or NTs based on Weibull and Gompertz distributions, respectively. The formulas are compared with the current exponential formula to examine their performance. The simulation results show that the sample size formula based on the Weibull distribution is the most robust among the three formulas in STs or NTs. We suggest that recognizing the appropriate distribution in advance is beneficial for proper project planning and that assuming a Weibull distributed survival time is most advantageous in STs or NTs.

**Keywords**:
Survival analysis; Sample size; Exponential distribution; Weibull distribution; Superiority trials; Non-inferiority trials

# 1. Introduction

A time-to-event endpoint is used as the primary endpoint in many studies such as those on oncology and cardiovascular disease. In the planning stage of a clinical trial employing this type of endpoint, determining an adequate sample size is one of the most fundamental steps. An appropriate sample size provides reasonable power to detect a clinically meaningful difference between treatment groups. Currently, sample size calculations usually assume that the failure time is exponentially distributed. Consider Quon's study [1], which examined whether the addition of weekly cisplatin to daily radiation therapy (RT) would improve survival in patients with unresectable squamous cell head-and-neck carcinoma. The formula based on the exponential distribution may have been used for the sample size calculation. However, the shape of the Kaplan-Meier curves was not consistent with the assumption that the event times are exponentially distributed, as was shown in Figure 2 in the original paper [1]. The hazard rates of both arms decreased along with the survival time. This issue may also arise in a clinical study incorporating a non-inferiority hypothesis. In other words, incorrect assumptions about the distribution of event times may result in an undesired sample size and power whether the study design uses superiority trials (STs) or non-inferiority trials (NTs).

A few derivations of sample size formulas for time-to-event data have been published that were based on distributions other than exponential [2-5]. Heo et al [2], for instance, derived a sample size formula that compared two groups of Weibull distributed survival times based on Schoenfeld and Richter [6]. Gope [3] proposed a sample size calculation method for comparing fatigue time following a Weibull distribution in technology studies. In addition, a simulation based method for calculating sample sizes for group sequential trials under a Weibull distribution was proposed by Jiang et al [4]. Lu et al [5] derived sample size formulas for a Weibull model in order to design a two-stage seamless adaptive trial under different hypotheses. In 2015, Wu [7] derived three sample size formulas based on a Weibull distribution and some researchers' work[6,8-10], for a randomized phase III clinical trial with general entry distribution. one of the claims made in this paper is that the Weibull distribution is superior to the exponential distribution because of the inclusion of an additional parameter. Heo's formula[2] is the same as Wu's first formula without the loss to follow-up, and Rubinstein's formula[11] was the same as Wu's first formula under the circumstances with the uniform entry, exponential survival distribution and loss to follow-up. All these articles under STs aimed at demonstrating the performance of the presented formula under different distributions, where no one discussed the discrepancy when the wrong distributions are hypothesized.

With respect to STs, Schoenfeld [8], Freedman [12], and Rubinstein's [11] formulas are frequently used, where Rubinstein's formula[11] has been used more frequently than the other two because of its consideration of accrual rate and censoring. Hence, the derived sample size formula with Weibull and Gompertz distribution for STs is planned to be based on Rubinstein's research.

In terms of sample size determination in NTs, some recently published methods assumed an exponential distribution [13-17]. In Rothmann's work [13], the focus was to examine issues involving retention of a pre-specified fraction of the control effect. Some design considerations were also mentioned in the paper, such as the formulation of the hypotheses, the statistical methodology and the interpretations of an active control non-inferiority trial. Sample size formulas for NTs as discussed by Chow et al. [15] and Crisp and Curtis [16], which extended the work of Lachin and Foulkes [18] from superiority hypothesis testing to non-inferiority, have been adopted in software packages such as nQuery 7.0, PASS 12.0, etc. The formula's power is biased in an unbalanced design study, and the bias increases as the two arms become more unbalanced or as the projected non-inferiority margin becomes farther from 1 [14]. Hence, Jung et al. [14] proposed a more accurate formula based on a non-inferiority log-rank test [15]. Subsequently, Jung and Chow [17] derived a generalized log-rank test and its sample size formula, which is more flexible (any survival distributions for two arms and any accrual pattern) and applicable for both superiority and non-inferiority inference. This sample size formula is identical to that of the log-rank test by Schoenfeld [8] (if the hazard ratio is set to 1 under the null hypothesis) and is also identical to that of Jung [14] (if the hazard ratio is 1 under the alternative hypothesis).

Commercially available software packages, including PASS, nQuery and EAST, can carry out the sample size based on the exponential distribution as well as a proportional hazard model. The same is true for standard textbooks [15,19], where sample size calculations under a Weibull model are not usually considered [7]. Because of the importance of the sample size, it is important to choose not only the methods of estimation but also the assumed distributions with caution. Although an exponential distribution may provide a reasonable approximation to the distribution of survival times over relatively short intervals, it typically does not adequately characterize the overall distribution of survival times because of its property of constant hazard over time [2]. Therefore, the Weibull and Gompertz distributions are considered in this paper. Two families of formulas based on the studies from Rubinstein [11] and Crisp and Curtis [16] are presented, and the robustness of three distributions in various situations are simulated.

The sample size formulas based on the three distributions and two hypotheses are shown in the next section. The setting and results of simulations are described in the *Simulations* and *Results* sections. The section entitled *Examples* gives two real clinical trials as examples. We then summarize our conclusions in the final section.

## 2. Methods

Consider a study with two treatment groups, one control ($x=0$) and one experimental ($x=1$). The effect size of the study when incorporating a time-to-event endpoint is expressed as a hazard ratio, which is defined as $\Delta = \lambda_1(t) / \lambda_0(t)$, where $\lambda_0(t)$ and $\lambda_1(t)$ are the hazard rates in the control and experimental group, respectively. If $\Delta$ is constant, the proportional hazard assumption is satisfied. $N_0$

and $N_1$ indicate the sample size of the control and experimental groups, respectively. Thus, the total sample size is $N = N_0 + N_1$. Only the balanced design is considered here, i.e., $N_0 = N_1$. The durations of enrollment and follow-up are denoted by $R$ and $T_f$, respectively, and the number of enrolled participants is assumed to have a Poisson distribution with parameter $N/R$. Suppose that our primary endpoint is the time to an undesirable event, such as death, progression of cancer, etc. So, the higher the hazard rate is, the worse the efficacy is.

## Formulas for STs

The hypotheses in the STs are usually set as $H_0: \Delta = 1$ versus $H_1: \Delta \neq 1$. The general form of the sample size formula from Rubinstein [11] is expressed as

$$N_0 = N_1 = \left(\frac{Z_{1-\alpha/2} + Z_{1-\beta}}{\ln(\Delta_1)}\right)^2 \left[E_0^{-1} + E_1^{-1}\right], \tag{1}$$

where $\Delta_1 (\Delta_1 > 1)$ is the hazard ratio under $H_1$, and $Z_\tau$ denotes the $(1-\tau)$ quantile of the standard normal distribution. This is the first sample size formula to consider the enrollment period and accrual rate, which is reflected in the expression of $E_0$ and $E_1$.

When the assumed survival distribution is exponential, i.e., the hazard function $h_x(t) = \lambda_x$, the expression for $E_x$ is written as

$$E_x = \frac{\lambda_x}{\lambda_x + \phi}[1 - \frac{e^{-T_f(\lambda_x + \phi)} - e^{-(T_f + R)(\lambda_x + \phi)}}{R(\lambda_x + \phi)}], \tag{2}$$

where $\lambda_x$ is the only parameter involved in the distribution of survival time, and $\phi$ denotes the hazard rate of identical exponential censoring distributions in both groups.

If a Weibull distribution is assumed to be used for describing the survival time where the hazard function is $h_x^W(t) = \lambda_x t^{k-1}$, $E_x$ is expressed as

$$E_x^W = \int_{T_f}^{T_f+R} \frac{1}{R} \int_0^z k\lambda_x u^{k-1} e^{-\lambda_x u^k} e^{-\phi_x u} du dz$$

$$= \frac{k}{R}\left( \int_0^{T_f} R\lambda_x u^{k-1} \exp(-\lambda_x u^k - \phi_x u)du + \int_{T_f}^{T_f+R} (T_f + R - u)\lambda_x u^{k-1} \exp(-\lambda_x u^k - \phi_x u)du \right), \quad (3)$$

where $\lambda_x$ is the scale parameter of the Weibull distribution in group $x(x=0,1)$. $k$ is the shape parameter of the Weibull distribution and is assumed to be identical in the two groups. Hence, the hazard ratio $\Delta = \frac{k\lambda_1 t^{k-1}}{k\lambda_0 t^{k-1}} = \frac{\lambda_1}{\lambda_0}$ is a constant.

When the distribution of survival times is assumed to be Gompertz, i.e., the hazard function $h_x^G(t) = \theta_x \exp(\alpha^G t)$, the $E_x$ in (1) is replaced by

$$E_x^G = \int_{T_f}^{T_f+R} \frac{1}{R} \int_0^z \theta_x \exp(\alpha^G u) \exp\{\theta_x/\alpha^G [1 - \exp(\alpha^G u)]\} \exp(-\phi_x u) du dz$$

$$= \frac{\theta_x}{R}\left( \int_0^{T_f} R \text{fun}(u) du + \int_{T_f}^{T_f+R} (T_f + R - u)\text{fun}(u) du \right), \quad (4)$$

where $\text{fun}(u) = \exp(\alpha^G u) \exp\{\theta_x/\alpha^G [1 - \exp(\alpha^G u)]\} \exp(-\phi_x u)$. $\alpha^G$ and $\theta_x$ are the shape parameter and scale parameter, respectively. The shape parameter is again assumed to be identical across the two groups in order to ensure that the proportional hazard assumption is met.

## Formulas for NTs

In NTs, the non-inferiority margin of the hazard ratio ($\tilde{\Delta}_0$) is one of the most important pre-specified values. If the observed hazard ratio exceeds $\tilde{\Delta}_0$, we declare the experimental group to be inferior to the control group; otherwise, the experimental group is non-inferior to the control group. So, given $\tilde{\Delta}_0$ in NTs, we intend to test $H_0: \Delta \geq \tilde{\Delta}_0$ against $H_1: \Delta < \tilde{\Delta}_0$. The hazard ratio under $H_1$ is indicated by $\tilde{\Delta}_1$ ($\tilde{\Delta}_1 < \tilde{\Delta}_0$). Under $H_1$, the efficacy of the experimental group is identical to (or possibly better than) that of the control group ($\tilde{\Delta}_1 \leq 1$) or slightly worse than that of the control group but within acceptable limits ($1 < \tilde{\Delta}_1 < \tilde{\Delta}_0$). To calculate the sample size, the general form of Crisp and Curtis's formula [16] for NTs is:

$$N_0 = N_1 = \left(\frac{Z_{1-\alpha/2} + Z_{1-\beta}}{\ln(\tilde{\Delta}_0) - \ln(\tilde{\Delta}_1)}\right)^2 \left[\tilde{E}_0^{-1} + \tilde{E}_1^{-1}\right]. \qquad (5)$$

The expression for $\tilde{E}_x$ in NTs for the exponential (Formula (2)), Weibull (Formula (3)) and Gompertz (Formula (4)) survival distributions is the same as that for the STs.

## Determination of study duration

It is difficult to directly derive the formula for study duration $T_f$, thus, we recommend using the numerical calculation (e.g. Newton-Raphson method) to solve

$$N_{setting} - N(T_f) = 0 \qquad (6)$$

for $T_f$, where the $N_{setting}$ is the pre-setting targeted number to enroll, The $N(T_f)$ is the function (1) or (5) of $T_f$ as the unknown parameter. To make (6) have a solution, $N_{setting}$ must satisfy

$$\lim_{T_f \to \infty} N(T_f) < N_{setting} < \text{ET}s \times \sum_{x=0}^{1} \left\{\frac{k}{R}\int_0^R (R-u)\lambda_x u^{k-1}\exp(-\lambda_x u^k - \phi_x u)du\right\}^{-1}$$

for exponential (k=1) and Weibull survival distribution and

$$\lim_{T_f \to \infty} N(T_f) < N_{setting} < \text{ET}s \times \sum_{x=0}^{1} \left\{\frac{\theta_x}{R}\int_0^R (R-u)\text{fun}(u)du\right\}^{-1},$$

where $\text{ET}s = \left(\frac{Z_{1-\alpha/2} + Z_{1-\beta}}{\ln(\Delta_1)}\right)^2$ under STs and $\text{ET}s = \left(\frac{Z_{1-\alpha/2} + Z_{1-\beta}}{\ln(\tilde{\Delta}_0) - \ln(\tilde{\Delta}_1)}\right)^2$ under NTs.

## Type of enrollment distribution

In the aforementioned formulas, the enrollment distribution is assumed to be uniform over interval [0, R], where the accrual rate is constant, and the number enrolled participants has a Poisson distribution. The distribution with a decreasing accrual rate presented in Lachin and Foulkes' study [18] is the truncated exponential distribution. It provides a great deal of flexibility in describing the accrual rates over a fixed period [20]. Any density function of the accrual distribution can be employed by replacing the $\frac{1}{R}$ in the integral in $E_x$. See Maki's paper [20] for other types of accrual distributions, such as increasing then constant, constant then decreasing, etc.

## 3. Simulations

Monte Carlo simulation studies were undertaken to verify the performance of sample size formulas when the survival time conformed to exponential, Weibull or Gompertz distributions in STs and NTs, respectively. While the simulations for STs and NTs use different sample size formulas and parameter settings, the generating processes for failure time and status are the same.

The independently observed event times $T_0$ are generated for the control and experimental groups under the assumed event rates, the hazard ratio ($\Delta_1$) and the censoring rate. For each subject, an enrollment time $r$ is generated from the uniform distribution over $[0, R]$, and a censoring time is generated from the exponential distribution with parameter $\phi$. As mentioned above, the subjects are treated as censored if the time-to-event is not the minimum of the time-to-event, censoring time or closure time minus enrollment time. Thus, the final observed time is $T = \min(T_0, C, T_f + R - r)$. Let $d$ indicate that the event either occurs ($d = 1$) or is censored ($d = 0$).

Then, 10000 independent trials were simulated for each sample size formula under each combination of parameters. Let type I error ($\alpha$) and power（$1 - \beta$） be 0.05 and 0.80, respectively. Different values for $T_f$ and $R$ are selected for satisfying Weibull and Gompertz simulations. When the survival time was Weibull distributed, the decreasing (k={0.5, 0.7}), constant (k=1.0) and increasing (k={1.2, 1.5}) hazard rates were considered, and the incidence intensity of the control group was set to low, intermediate or high ($\lambda_0 = \{0.1, 0.3, 0.5, 1.0, 1.2\}$). In the simulations in which the failure time distribution was Gompertz, the shape and scale parameters were set to $\alpha^G = \{0.5, 0.7, 1.0, 1.2, 1.5\}$ and $\theta_0 = \{0.1, 0.3, 0.5, 1.0, 1.2\}$, respectively. Random censoring absent ($\phi = 0$) and present ($\phi = 0.2$) were also considered. The parameters used in the sample size formulas were estimated using the averaged coefficients of parametric regression model of corresponding distribution, which was fitted to 20 trials of 50 subjects' failure time generated by the real survival distribution with preseted parameters. Only the balanced design is considered in this section, i.e. $N_0 = N_1 = N/2$

Additionally, the detail simulating process is illustrated. Firstly, 20 trials of 50 subjects with a preset survival distribution, say Weibull, were generated. Then, exponential, Weibull and Gompertz regression models were fitted, and the averaged regressed coefficients were used for calculating the parameters utilized in each sample size formula. After the calculation of 3 sample sizes under this circumstance, 10000 samples within each sample size group were taken from the preset distribution. As a result, the empirical powers of each sample size formula in each parameter combination of Weibull situation can be observed.

For STs, the rejection of $H_0 : \Delta_0 = 1$ by a two-sided Wald test of a Cox proportional hazards regression model yields the conclusion that the hazard rates of the two groups are not equal. The true hazard ratio $\Delta_1 = \left\{ \frac{1}{1.2}, \frac{1}{1.5} \right\}$ is then used to determine the sample size. For NTs, the non-inferiority margin $\tilde{\Delta}_0$ was 1.2 and 1.5 while the true hazard ratio $\tilde{\Delta}_1$ was 1 to imply the same efficacy for the two groups. If the upper limit of the two-tailed 95% confidence interval (for the hazard ratio from a Cox proportional hazards regression model) is less than the non-inferiority margin, then non-inferiority is concluded.

The parameter combinations for the Weibull and Gompertz failure time distributions are shown in Tables 1 and 2, respectively. The statistical package R 2.15.2 was used to perform simulations under different survival distributions and to generate tables and graphs.

## 4. Results

In situations having the same $k$, $\lambda_0$ and $\phi$ between STs and NTs, the required events are equal if $|\log(\Delta_0 / \Delta_1)|$ is identical, while the final sample size differs

because of different hypotheses (different hazard rate in the experimental group). However, the key conclusion is analogous.

The simulations and results for estimated study duration are similar as for the sample size since the formula of study duration is resulted from the sample size formula directly. The larger sample size needed with different survival distribution, and the larger study duration calculated. If the sample size is overestimated, the study duration is overestimated as well, and vice versa.

## Results of Superiority Trials

The empirical power function assuming Weibull survival time in STs is shown in Figure 1 and Table A1. When the survival time has a Weibull distribution, the power using the Weibull formula is very close to the specified power, which is to be expected.

The powers of both the exponential formula and Gompertz formula are heavily influenced by the shape parameter and censoring rate. In cases where the Weibull distribution has a decreasing hazard function with survival time ($k<1$) (Figure 1-A, 1-B, 1-F, 1-G), the empirical power of the exponential and Gompertz formula decreases from 1.0 to around 0.8 with increasing $\lambda_0$ and decreasing censoring. The sample size from exponential is larger than that from the Gompertz. When $k=1$, the exponential formula is identical to the Weibull formula (Figure 1-C,1-H). The power of Gompertz formula is around 0.8 as well. When k is larger than 1.0 with random censoring (Figure 1-D, 1-E, 1-I, 1-J), the power of the both formulas are a little lower than 0.8 with high censoring rate, and approaching 0.8 with an increasing $\lambda_0$ and a decreasing censoring rate.

The results from simulation of the Gompertz survival time (Figure 2) show that the power calculated using the Weibull and exponential formula are not smaller than that of the Gompertz in presence of high censoring rate. The largest difference between Weibull and Gompertz is 0.225 when the censoring rate reaches 63.2% where the largest difference between exponential and Gompertz is 0.420. Along with the censoring rate decreasing, the power lines of exponential and Weibull cross the power line of Gompertz. The power of the other formula are larger than 0.8. But Weibull formula is closer to the Gompertz formula than exponential formula, where the power's differences between Weibull and Gompertz are less than 0.04. The details are listed in Table A2.

**Results of non-inferiority trials**

The empirical power functions using Weibull survival time in NTs are shown in Figure 3, and the details are listed in Table A3.

The trends for the Gompertz distribution in the NTs are almost the same as those for STs. Power increases from approximately 0.6 to over 0.9 along with increasing $\lambda_0$, and the distances between the Gompertz and Weibull lines increase as the censoring rate increases. However, the trends for the exponential formula in the NTs differ from those in STs.

For $k<1$ (Figures 3-A, 3-B) with random censoring ($\phi=0.2$), the exponential formula may overestimate the sample size; in addition, the largest empirical power for the exponential distribution is greater than 0.95, substantially larger than the pre-specified 0.80. So, using the exponential formula in this scenario may yield an unnecessarily large sample. In the case of $k=1$ (Figure 3-C), the formulas under the exponential and Weibull distributions are identical, and power is approximately 0.8. When $k>1$ (Figure 3-D, 3-E), the exponential formula is closer to the necessary sample size with an increasing hazard function of survival time.

In Figure 4, the results are similar to those from STs. The Weibull formula approaches the Gompertz formula. The details are listed in Table A4.

Overall, when the survival time is Weibull distributed, sample sizes from the exponential and Gompertz formulas do not estimate the necessary sample size correctly. When the survival time is Gompertz distributed, the exponential sample size formula may underestimate the sample size needed, whereas the Weibull sample size formula is closer to the Gompertz formula and provides a reasonable estimation of the required sample size.

## 5. Examples

### Example 1: Superiority clinical trial

The sample size formula for STs is demonstrated using a phase III study of radiation therapy with ($x=1$) or without ($x=0$) cis-platinum in patients with unresectable squamous or undifferentiated carcinoma of the head and neck [1]. The major endpoint is failure-free survival measured from the date of randomization to the first evidence of progression, relapse or death. Because the Weibull formula has been shown to be more robust than the others, only the Weibull distribution sample size is illustrated in this example.

In this study, 300 patients (150 patients in each arm) were needed to detect a 50% increase in the median survival between the two arms with 80% power and an overall Type I error of 0.05 by a two-sided test. So, the hazard ratio under $H_1$ was assumed to be 1/1.5=0.667. An accrual rate of 75 patients/year was expected and therefore the participants were enrolled in 48 months (300/75*12). The maximum follow up time was 13 years from the survival plots illustrated (Fig 2-3 in the original paper).

The relationship between the median survival time and the Weibull hazard rate is $\lambda = -\ln(\text{survival rate})/(k\text{th power of survival time at the rate})$. While the expected median survival of 13 months in the control arm (the researchers decide in the original paper), the exponential rates are $\lambda_0 = -\ln(0.5)/13^1 = 0.053$ and are $\lambda_1 = 0.080$ in the case of $k=1$. The calculated sample sizes for several situations are shown in Table 3.

When the survival distribution is exponential, the sample size needed per group is 104 if there is no censoring and 193 if there is exponentially distributed censoring with a parameter of 0.05, which needs almost 90 patients larger than the case of no censoring. The calculated result of $k=1.0$ is very close to that for $k=1.5$ since the increment caused by the short research period is close to 0.

## Example 2: Non-inferiority clinical trial

The sample size considerations for NTs are illustrated by a randomized, open-label, phase III, parallel clinical study [21]. The sample size was conducted to compare the efficacy on the patients with gastric cancer of Capecitabine/cisplatin (XP, experimental arm $x=1$) versus 5-fluorouracil/cisplatin (FP, standard/control arm, $x=0$). The primary outcome of the trial was progression-free survival (PFS), measured as time from randomization to the first date of documented disease progression or death. A total of 316 patients were randomized 1:1 to receive the XP and FP.

The primary analysis of PFS was of the non-inferiority of XP to FP, as measured by the hazard ratio $\Delta_{\text{XP/HP}}$ with a non-inferiority margin of 1.40. $H_0$ was rejected if the upper limit of the 95% two-tailed confidence interval of the hazard ratio calculated using the Cox proportional hazards model was less than the non-inferiority margin.

The median PFS was 5.6 months and 5.0 months for XP and FP, respectively. The exponential hazard rate can be derived as $\lambda_0 = 0.139$. The accrual interval was 22 months, and the maximum follow-up time was approximately 24 months according to their survival plots. Because there is no description of the censoring, it appears in the paper's figures that there was no random censoring in the perprotocol population.

The parameter combinations with random censoring and without random censoring are considered. Under an exponential assumption, 141 patients and 190 patients are required without and with random censoring, respectively. In the case of decreasing hazard rates (i.e., $k = 0.5$), the sample size that is needed is larger than that of the constant hazard rate because of the low incidence rate. If the hazard rate is increasing, the event incidence rate is slightly higher so that the numbers of required subjects decrease to 140 and 183.

## 6. Conclusions

In this paper, we presented two families of sample size formulas for exponential, Weibull and Gompertz distributions in superiority trials and non-inferiority clinical trials. A substantial number of simulations were conducted to investigate the robustness of these formulas. Unbalanced designs and disproportional hazards must be worked out in future studies.

As shown in this paper, some situations may result in a waste of resources or in an underestimation of the sample size. The Weibull sample size formula worked relatively well in the simulated scenarios, in which the proportional hazard assumption held. Therefore, we suggest that recognizing the appropriate distribution in advance is very helpful for planning the project appropriately, and it is better to assume a Weibull distributed survival time in superiority trials or non-inferiority clinical trials unless there is clear evidence that the survival time has some other distribution.

**Acknowledgments**

This work was supported by the National Natural Science Foundation of China (81673268) and the Natural Science Foundation of Guangdong Province, China (2017A030313812).

**Table 1. The parameter set for simulating trials when failure time is Weibull distributed**

| Parameters | Values for STs | Values for NTs |
|---|---|---|
| $k$ | 0.5, 0.7, 1.0, 1.2, 1.5 | 0.5, 0.7, 1.0, 1.5, 2.0 |
| $\lambda_0$ | 0.1, 0.3, 0.5, 1.0, 1.2 | 0.5, 0.7, 0.8, 1.0, 2.0 |
| $\phi$ | 0, 0.2 | 0, 0.2 |
| $(T_f, R)$ | (6,2) | (12,2) |
| $(\Delta_1, \Delta_0)$ or $(\tilde{\Delta}_1, \tilde{\Delta}_0)$ | (1/1.2,1.0), (1/1.5,1.0) | (1.0,1.2), (1.0,1.5) |

Notes: $\lambda_1 = \Delta_1 \cdot \lambda_0$; $\alpha = 0.05$; Power = 0.8

**Table 2. The parameter set for simulating trials when failure time is Gompertz distributed**

| Parameters | Values for STs | Values for NTs |
|---|---|---|
| $\alpha^G$ | 0.5, 0.7, 1.0, 1.2, 1.5 | 0.05, 0.1, 0.5, 1.0 |
| $\theta_0$ | 0.1, 0.3, 0.5, 1.0, 1.2 | 0.05, 0.1, 0.5, 1.0 |
| $\phi$ | 0, 0.2 | 0, 0.2 |
| $(T_f, R)$ | (2,1) | (12,2) |
| $(\Delta_1, \Delta_0)$ or $(\tilde{\Delta}_1, \tilde{\Delta}_0)$ | (1/1.2,1.0), (1/1.5,1.0) | (1.0,1.2), (1.0,1.5) |

Notes: $\lambda_1 = \Delta_1 \cdot \lambda_0$; $\alpha = 0.05$; Power = 0.8

**Table 3. Sample sizes needed for different parameter combinations assumed in Example 1**

| $k$ | $\lambda_0$ | $\lambda_1$ | $\Delta_1$ | $\phi$ | N(per group) |
|---|---|---|---|---|---|
| 0.5 | 0.192 | 0.288 | 0.667 | 0 | 104 |
|  |  |  |  | 0.05 | 193 |
| 1.0 | 0.053 | 0.080 | 0.667 | 0 | 97 |
|  |  |  |  | 0.05 | 187 |
| 1.5 | 0.015 | 0.022 | 0.667 | 0 | 97 |
|  |  |  |  | 0.05 | 183 |

Notes: $\Delta_0 = 1$, $T_f = 156$, $R = 48$, $\alpha = 0.05$, power=0.8; number of events=96

**Table 4. Sample sizes needed for different parameter combinations assumed in Example 2**

| $k$ | $\lambda_0$ | $\tilde{\Delta}_0$ | $\phi$ | N(per group) |
|---|---|---|---|---|
| 0.5 | 0.310 | 1.40 | 0 | 167 |
|  |  |  | 0.05 | 218 |
| 1.0 | 0.139 | 1.40 | 0 | 141 |
|  |  |  | 0.05 | 190 |
| 1.5 | 0.062 | 1.40 | 0 | 140 |
|  |  |  | 0.05 | 183 |

Notes: $\tilde{\Delta}_1 = 1$, $T_f = 24$, $R = 22$, $\alpha = 0.05$, power=0.8; number of events=139

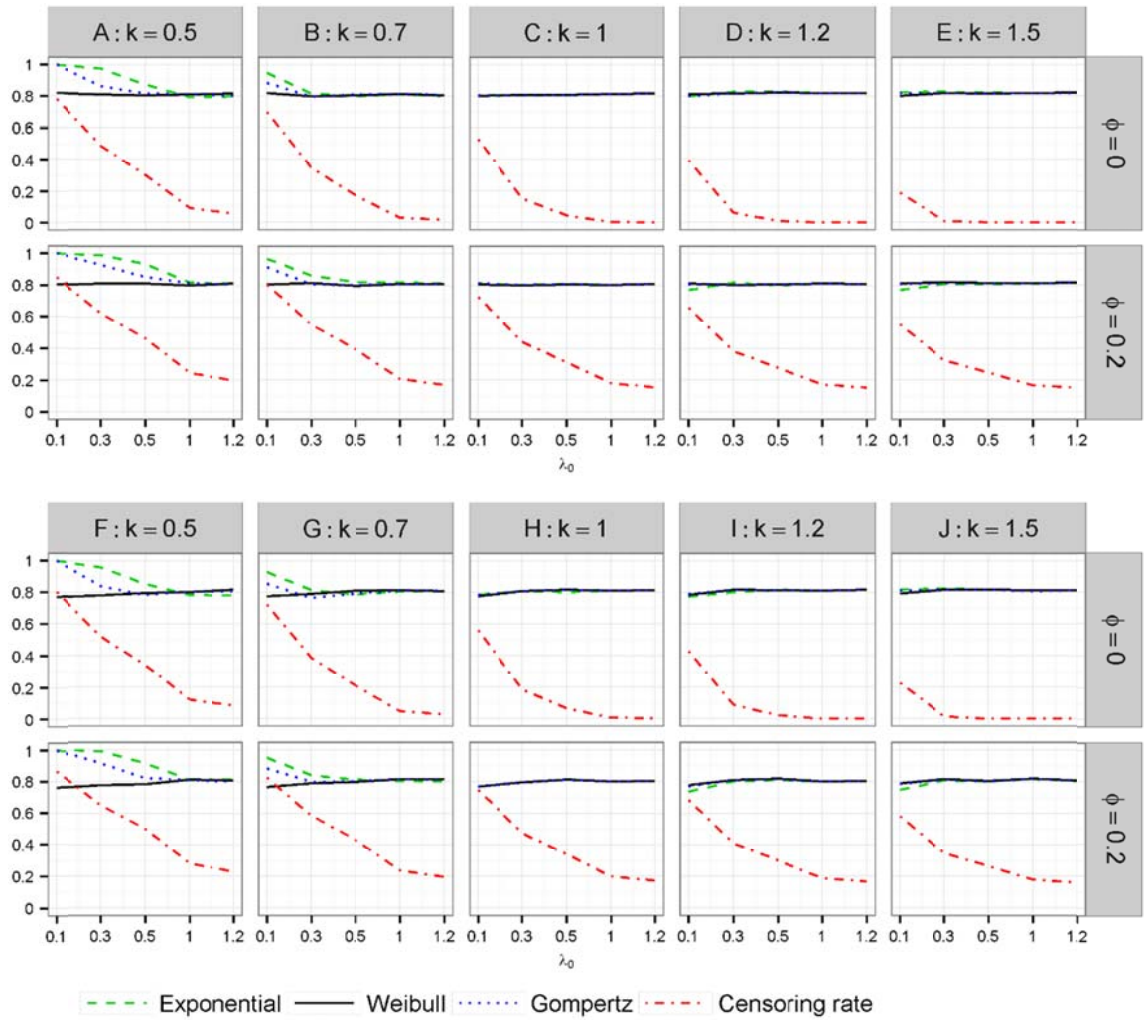

**Figure 1. The power using the sample size formulas from three distributions when the failure time is Weibull distributed in STs.**

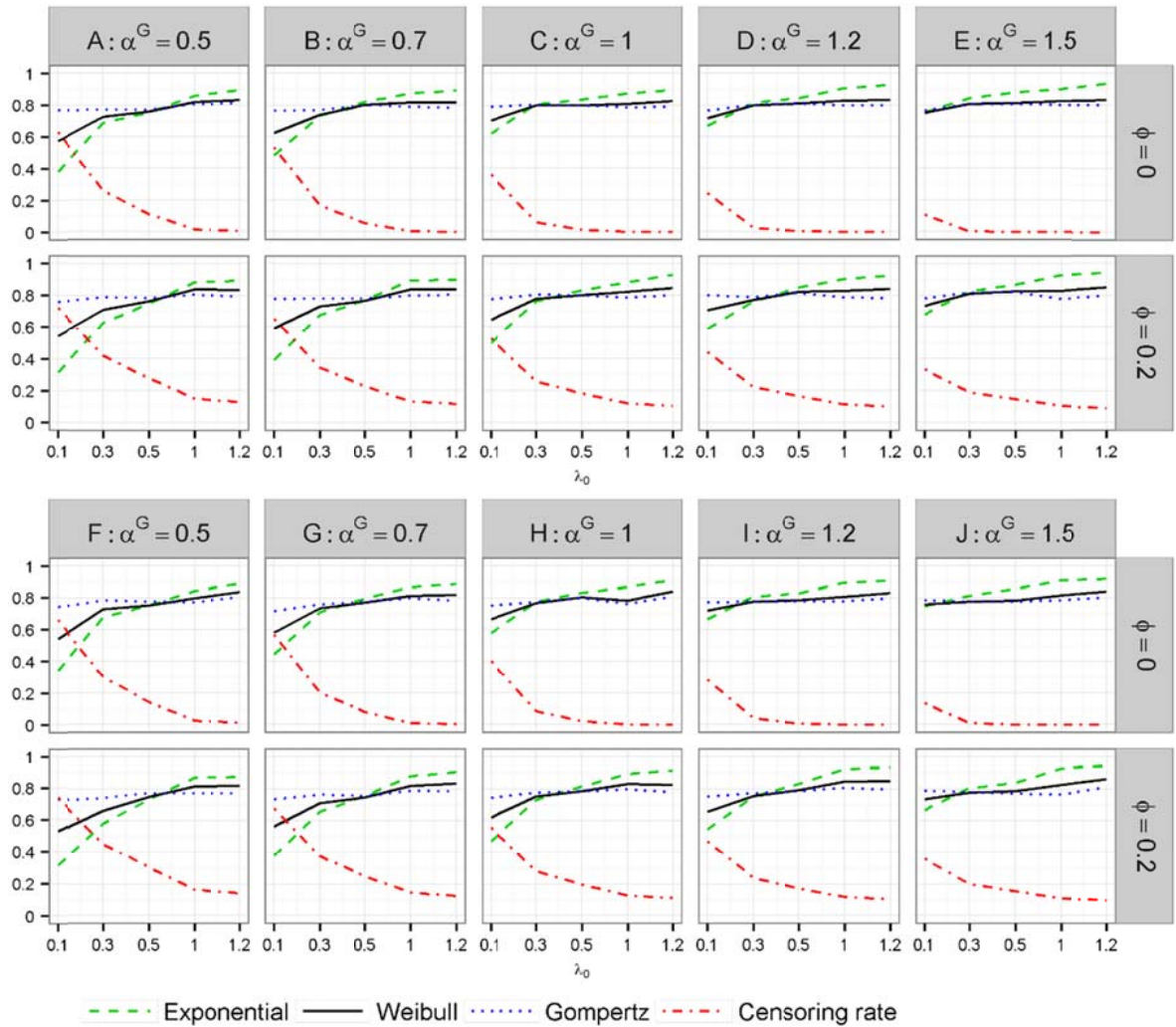

**Figure 2. The power using the sample size formulas from three distributions when the failure time is Gompertz distributed in STs**

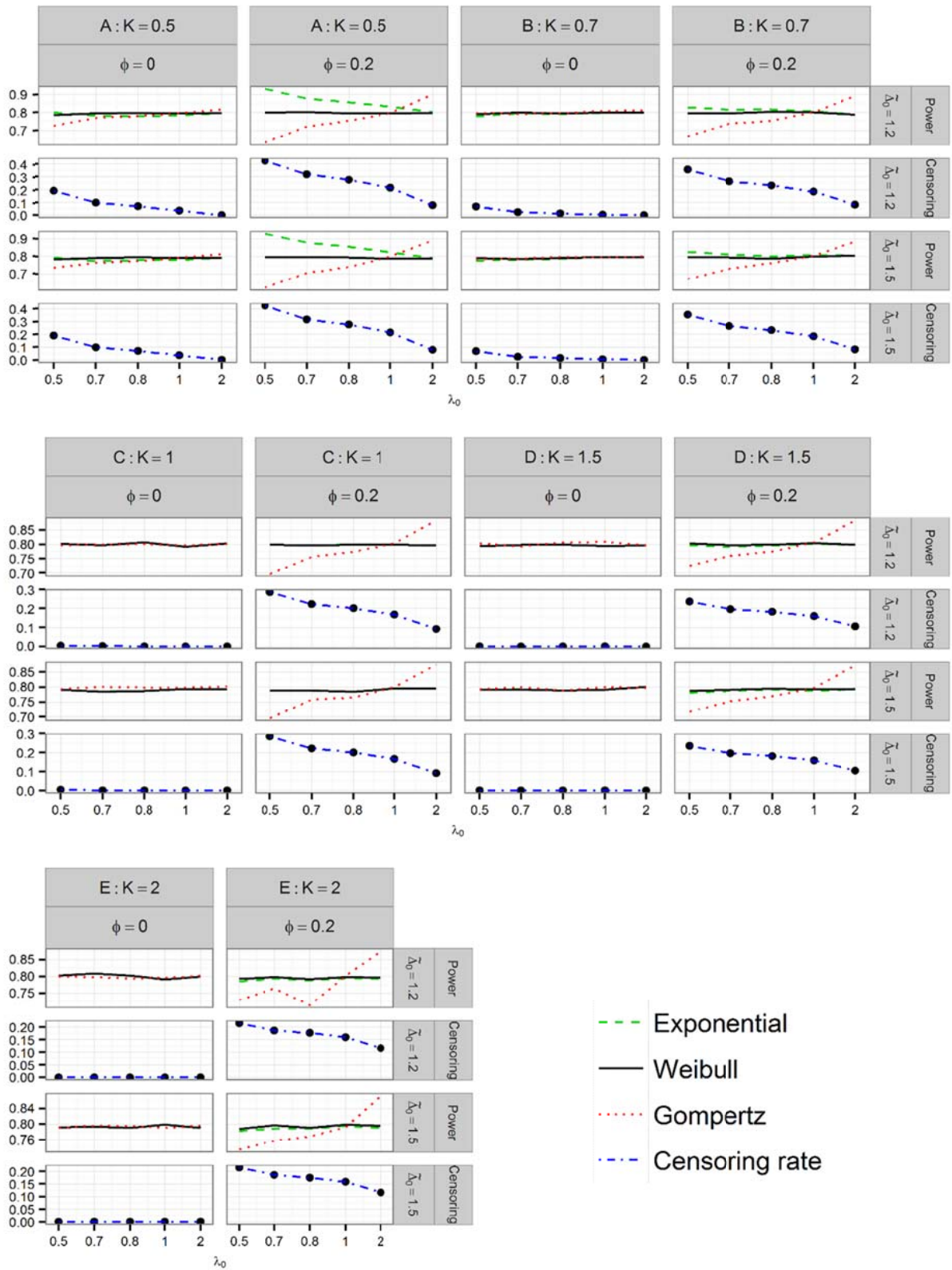

Figure 3. The power using the sample size formulas from three distributions when the failure time is Weibull distributed in NTs.

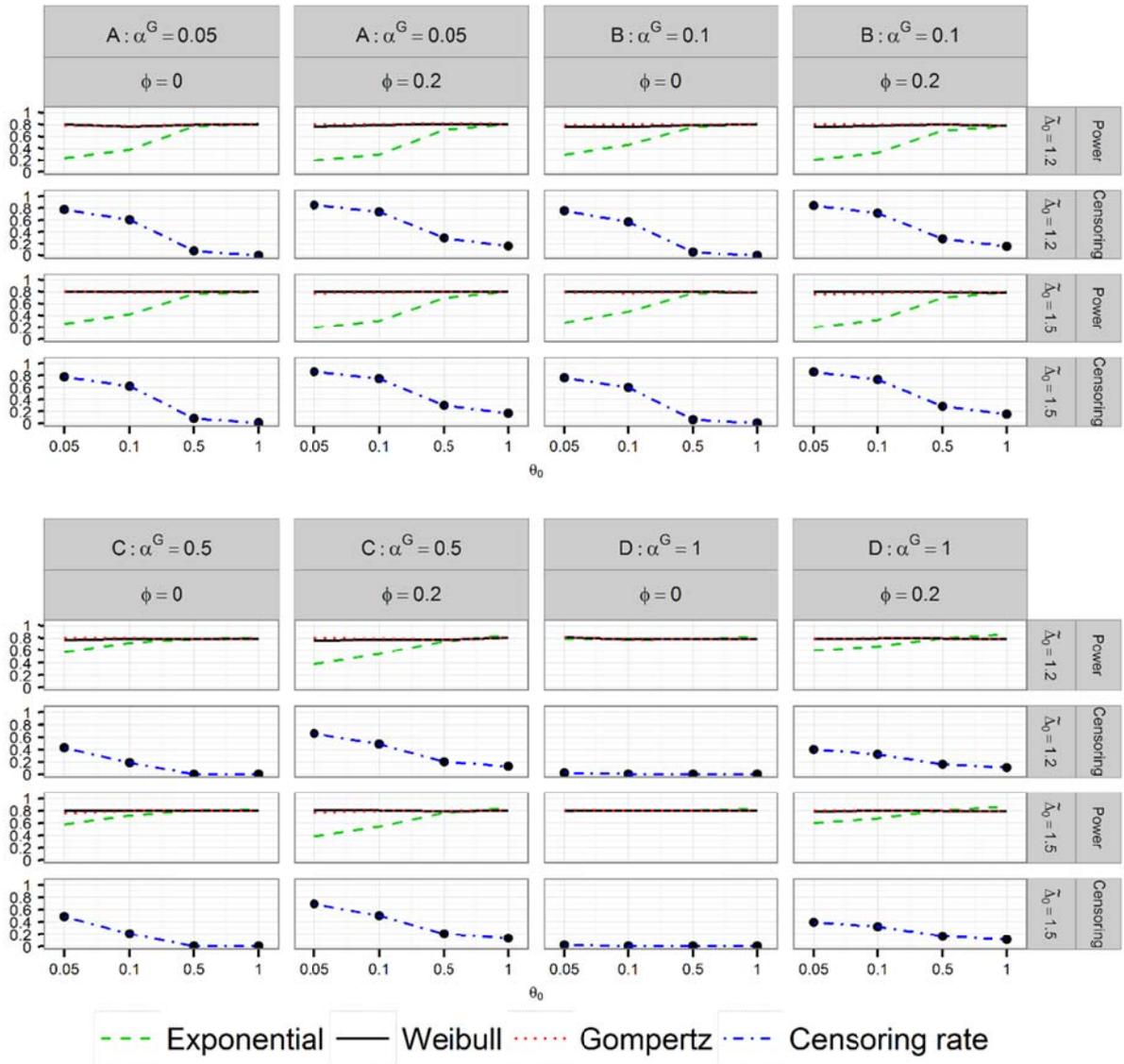

**Figure 4.** The power using the sample size formulas from three distributions when the failure time is Gompertz distributed in NTs